\documentclass[preprint,aps,superscriptaddress]{revtex4}

\usepackage[dvipdfmx]{graphicx}
\usepackage{dcolumn}
\usepackage{bm}
\usepackage{color}

\begin{document}

\preprint{APS/123-QED}

\title{Charge correlation in V$_2$OPO$_4$ probed by hard x-ray photoemission spectroscopy}

\author{Kota~Murota}
\affiliation{Department of Applied Physics, Waseda University, Shinjuku, Tokyo 169-8555, Japan}
\author{Elise~Pachoud}
\affiliation{Centre for Science at Extreme Conditions and School of Chemistry, University of Edinburgh, Edinburgh EH9 3FD, United Kingdom}
\author{J. Paul Attfield}
\affiliation{Centre for Science at Extreme Conditions and School of Chemistry, University of Edinburgh, Edinburgh EH9 3FD, United Kingdom}
\author{Robert~Glaum}
\affiliation{Institut f{\rm $\ddot{u}$}r Anorganische Chemie, Universit{\rm $\ddot{a}$}t Bonn, D-53012 Bonn, Germany}
\author{Tatsunori~Yasuda}
\affiliation{Department of Applied Physics, Waseda University, Shinjuku, Tokyo 169-8555, Japan}
\author{Daiki~Ootsuki}    
\affiliation{Department of Interdisciplinary Environment, Kyoto University, Sakyo, Kyoto 606-8501, Japan}
\author{Yasumasa~Takagi}    
\affiliation{Japan Synchrotron Radiation Research Institute, Sayo, Hyogo 679-5198, Japan}
\author{Akira~Yasui}   
\affiliation{Japan Synchrotron Radiation Research Institute, Sayo, Hyogo 679-5198, Japan}
\author{Daniel~I.~Khomskii}
\affiliation{II Physikalisches Institut, Universit{\rm $\ddot{a}$}t zu K{\rm $\ddot{o}$}ln, 50937 K{\rm $\ddot{o}$}ln, Germany}
\author{Takashi Mizokawa}
\affiliation{Department of Applied Physics, Waseda University, Shinjuku, Tokyo 169-8555, Japan}

\date{\today}

\begin{abstract}
Electronic properties of V$_2$OPO$_4$ have been investigated by means of hard x-ray photoemission spectroscopy (HAXPES) and subsequent theoretical calculations. The V 1$s$ and 2$p$ HAXPES spectra are consistent with the charge ordering of V$^{2+}$ and V$^{3+}$. The binding energy difference between the V$^{2+}$ and V$^{3+}$ components is unexpectedly large indicating large bonding-antibonding splitting between them in the final states of core level photoemission. The V 1$s$ HAXPES spectrum exhibits a charge transfer satellite which can be analyzed by configuration interaction calculations on a V$_2$O$_9$ cluster. The V 3$d$ spectral weight near the Fermi level is assigned to the 3$d$ $t_{2g}$ orbitals of the V$^{2+}$ site. The broad V 3$d$ spectral distribution is consistent with the strong hybridization between V$^{2+}$ and V$^{3+}$ in the ground state. The core level and valence band HAXPES results indicate substantial charge transfer from the V$^{2+}$ site to the V$^{3+}$ site.
\end{abstract}

\pacs{   }
\maketitle

\newpage

\section{Introduction}
Electronic properties of transition-metal compounds are governed by Coulomb repulsion between $d$ electrons and hybridization between ligand and $d$ electrons \cite{Imada1998, Khomskii2014}. Among various transition-metal compounds, V oxides have been playing important roles in understanding of physics of metal-insulator transitions (MITs). In V$^{3+}$ with octahedral coordination, triply degenerate $t_{2g}$ orbitals accommodate two electrons, and their orbital polarization is essential for their MIT and magnetism. For example, V$_2$O$_3$ with corundum structure exhibits a MIT which is accompanied by magnetic and orbital orderings \cite{McWhan1973,Sawatzky1979}. In the face sharing V-V bond of V$_2$O$_3$, $a_{1g}e_{g}^{\pi}$ and $e_{g}^{\pi}e_{g}^{\pi}$ configurations are realized and provide ferromagnetic coupling of V spins \cite{Park2000}. Here, the $a_{1g}$ orbital has $3z^2-r^2$ symmetry when the $z$ axis is perpendicular to the shared face. Another example is LiVO$_2$ with triangular lattice in which the two $t_{2g}$ electrons per V site form single bonds in two different directions and help trimerization of three V sites \cite{Reuter1962,Kikuchi1991,Pen1997}. The trimerization and the face sharing bond coexist in BaV$_{10}$O$_{15}$ \cite{Kajita2010,Takubo2012}. A HAXPES study on BaV$_{10}$O$_{15}$ indicated that, for V$^{2+}$ and V$^{3+}$ located in the face sharing V sites, the binding energy difference is unexpectedly large \cite{Yoshino2017} compared to that between V$^{3+}$ and V$^{4+}$ \cite{Maiti2000,Takubo2006,Wadati2008,Suga2010a} or between V$^{4+}$ and V$^{5+}$ \cite{Konstantinovic2005,Morita2010,Suga2010b,Chen2014}. The mixed valence of V$^{2+}$ and V$^{3+}$ is rather rare and their electronic properties are less explored compared to the mixed valence of V$^{3+}$ and V$^{4+}$  and that of V$^{4+}$ and V$^{5+}$.

Very recently, Pachoud {\it et al.} have reported negative thermal expansion in V$_2$OPO$_4$ which is driven by charge transfer between V$^{2+}$ and V$^{3+}$ \cite{Pachoud2018, Pachoud2020}. V$_2$OPO$_4$ consists of face and corner sharing VO$_6$ octahedra as shown in Fig. 1. Below 605 K, the V$^{2+}$/V$^{3+}$ charge ordering along the face sharing chain is accompanied by monoclinic lattice distortion \cite{Pachoud2018}. The V$^{2+}$ and V$^{3+}$ spins are ferrimagnetically ordered below 165 K. The magnetic susceptibility above 165 K exhibits a Curie-Weiss behavior with paramagnetic moment of 1.61 $\mu_B$ per V$_2$OPO$_4$ unit, which is reduced from the ideal values for V$^{2+}$ and V$^{3+}$ spins. The reduction of the paramagnetic moment suggests strong hybridization between the V$^{2+}$ and V$^{3+}$ sites along the face sharing bond in which their spins are antiferromagnetically correlated. In addition, relationship between the magnetic order and V 3$d$ orbital order has been investigated by x-ray absorption spectroscopy \cite{Murota2020}. Although the strong hybridization between the V$^{2+}$ and V$^{3+}$ has been suggested to play essential roles in this unique system, it is not elucidated yet by means of photoemission spectroscopy. In the present work, we study electronic structure of V$_2$OPO$_4$ by means of hard x-ray photoemission spectroscopy (HAXPES) and subsequent theoretical calculations in order to understand the origin of unique structural and magnetic behaviors.  

\begin{figure}
\begin{center}
\includegraphics[width=10cm]{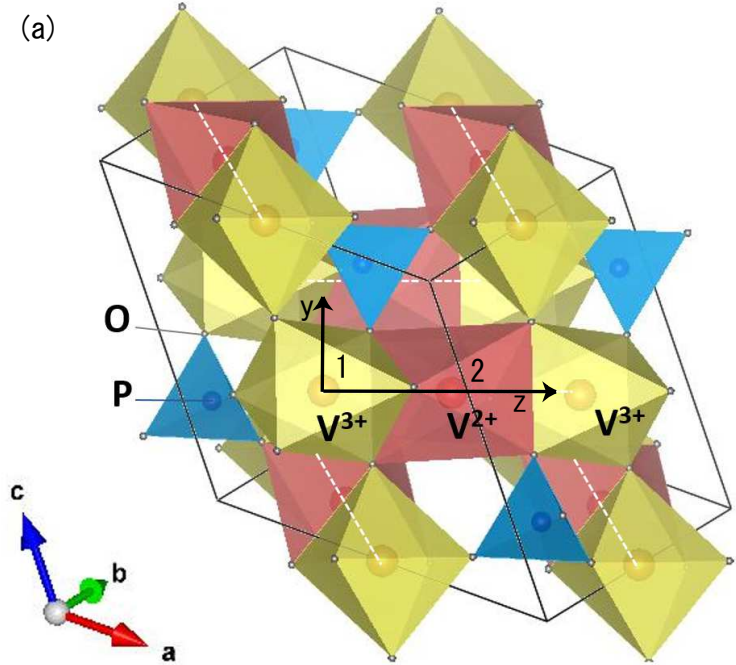}
\caption{
(Color online) 
Crystal structure of V$_2$OPO$_4$. The dashed lines indicate face sharing chains of the VO$_6$ octahedra. The $z$ axis is perpendicular to the shared face.
}
\end{center}
\end{figure}

\section{Methods}
The single crystals of V$_{2}$OPO$_{4}$ were grown as reported in the literatures \cite{Glaum1989,Pachoud2018}. HAXPES measurements were performed at BL47XU of SPring8 \cite{Ikenaga2013,Ikenaga2018}. Some of the crystals were cleaved {\it in situ} at 300 K under ultrahigh vacuum of 10$^{-6}$ Pa for measurements. Others were prepared {\it ex situ} in the air and then introduced to the vacuum for measurements. The photon energy was set to 7940 eV and the photoelectrons were collected and analyzed by VG Scienta R4000-10kV. The pass energy was set to 200 eV and the total energy resolution was about 270 meV. The binding energy was calibrated using the Fermi edge of Au reference. 

The V 1$s$ HAXPES spectrum was analyzed by the V$_2$O$_9$ cluster model calculations in which two VO$_6$ octahedra share its face. Five V 3$d$ electrons are accommodated in the cluster and the charge transferred configurations are considered. The ground state is given by a linear combination of $d_1^2d_2^3$, $d_1^3d_2^2$, and $d_1^3d_2^3L$ configurations in which $d_1$ and $d_2$ correspond to the V 3$d$ orbitals of the V$^{3+}$ and V$^{2+}$ sites, respectively. $L$ represents a hole in 27 ligand orbitals constructed from O 2$p$. The energy difference between the $d_1^2d_2^3$ and $d_1^3d_2^2$ is expressed as $E_G$ and is deduced to be 1.5 eV from the LDA+$U$ calculations. The energy difference between the $d_1^3d_2^3L$ configuration and the $d_1^2d_2^3$ corresponds to the O 2$p$ to V 3$d$ charge transfer energy $\Delta$ for V$^{3+}$ which is one of the three adjustable parameters \cite{Bocquet1996,Mizokawa1996}. The transfer integrals are parameterized by Slater-Koster parameters ($pd\sigma$) and ($pd\pi$) with ($pd\pi$)/($pd\sigma$)=-0.45 for the V 3$d$ and O 2$p$ orbitals. ($pd\sigma$) is one of the three adjustable parameters. The transfer integrals between the O 2$p$ orbitals are described by ($pp\sigma$) and ($pp\pi$) which are fixed to -0.40 eV and 0.10 eV, respectively. The transfer integral between the V 3$d$ $a_{1g}$ orbitals is given by ($dd\sigma$) which is fixed to -0.50 eV considering the V-V bond length of 2.68 ${\rm \AA}$ \cite{Takubo2006}. Variations of ($pp\sigma$), ($pp\pi$), and ($dd\sigma$) values within their chemical trends do not affect the main conclusion. The Coulomb interaction between the V 1$s$ and V 3$d$ electron $U_c$ is one of the three adjustable parameters. The exchange interaction between the V 3$d$ electron and that between the V 1$s$ and V 3$d$ electrons are set to 0.5 eV. 

LDA, LDA+$U$, and GGA+$U$ calculations were performed using QUANTUM ESPRESSO 5.30 \cite{QE1,QE2}. We employed pseudopotentials of V.pz-spnl-kjpaw\_psl.1.0.0.UPF, O.pz-n-kjpaw\_psl.0.1.UPF, and P.pz-n-kjpaw\_psl.0.1.UPF for LDA,
and those of V.pbesol-spnl-kjpaw\_psl.1.0.0.UPF, O.pbesol-n-kjpaw\_psl.1.0.0.UPF, and P.pbesol-n-kjpaw\_psl.1.0.0.UPF for GGA. $U$ was set to 3.5 eV or 5.0 eV. $J$ was set to 0.5 eV. Cutoff energy was set to 30 Ry \cite{Murota2020}.

\section{Results and Discussion}
Figure 2(a) shows V 1$s$ HAXPES spectra taken at room temperature for the clean and oxidized surfaces. The V 1$s$ spectrum from the surface exposed to the air is dominated by V$^{3+}$ component as shown in the lower panel of Fig. 2(b). The clean surface obtained by {\it in situ} cleaving exhibits V$^{2+}$ component on the lower binding energy side in addition to the V$^{3+}$ peak. Considering the mixed valence and the exchange splitting between the V 1$s$ core hole spin and the V 3$d$ spin, the V 1$s$ spectrum can be decomposed into exchange split V$^{2+}$ and V$^{3+}$ components by using four Gaussians as shown in the upper panel of Fig. 2(b). The ratio of V$^{2+}$ : V$^{3+}$ is estimated to be 4 : 3 for the spectrum of clean surface. Most probably, the spectral weight of V$^{3+}$ is reduced due to charge transfer from the V$^{2+}$ to V$^{3+}$ sites by the attractive force from the core hole. The binding energy difference between the V$^{2+}$ and V$^{3+}$ components is about 2.5 eV. This value is comparable to that reported for BaV$_{10}$O$_{15}$ and is almost twice the typical values for the difference of V$^{3+}$ and V$^{4+}$ \cite{Maiti2000,Takubo2006,Wadati2008,Suga2010a} or of V$^{4+}$ and V$^{5+}$ \cite{Konstantinovic2005,Morita2010,Suga2010b,Chen2014}. The large energy separation between V$^{2+}$ and V$^{3+}$ can be assigned to the effect of bonding-antibonding splitting between them in the final states of core level photoemission. In addition, a broad satellite structure is observed on the higher binding energy region. The satellite structure can be assigned to charge transfer from O 2$p$ to V 3$d$. The energy difference between the satellite and the main peaks is about 10-15 eV, consistent with typical charge transfer energy ($\sim$ 6-9 eV) and hybridization (5-6 eV) between the O 2$p$ and V 3$d$ orbitals for V$^{2+}$ and V$^{3+}$ oxides \cite{Bocquet1992,Khomskii2014}. It should be noted that the hybridization term is given by the transfer integral multiplied by the factor of $\sqrt{N}$ ($N$ is number of unoccupied 3$d$ orbitals).

\begin{figure}
\begin{center} 
\includegraphics[width=10cm]{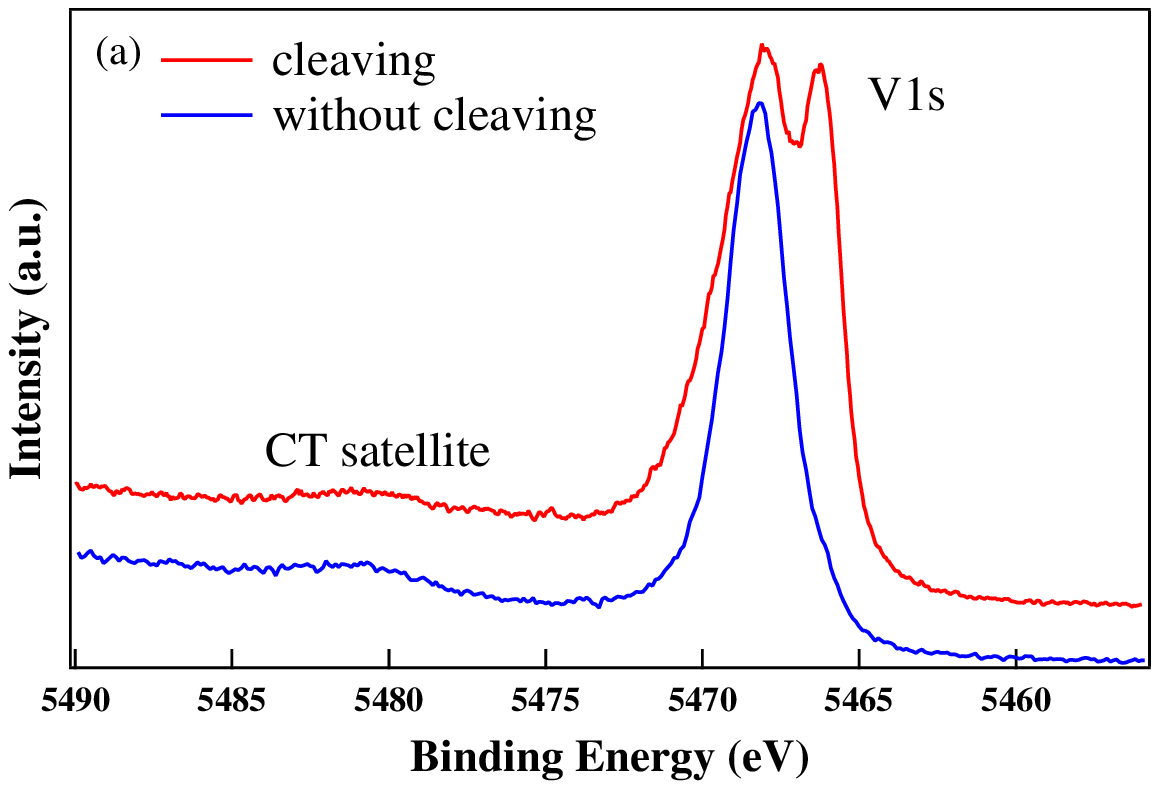}
\includegraphics[width=10cm]{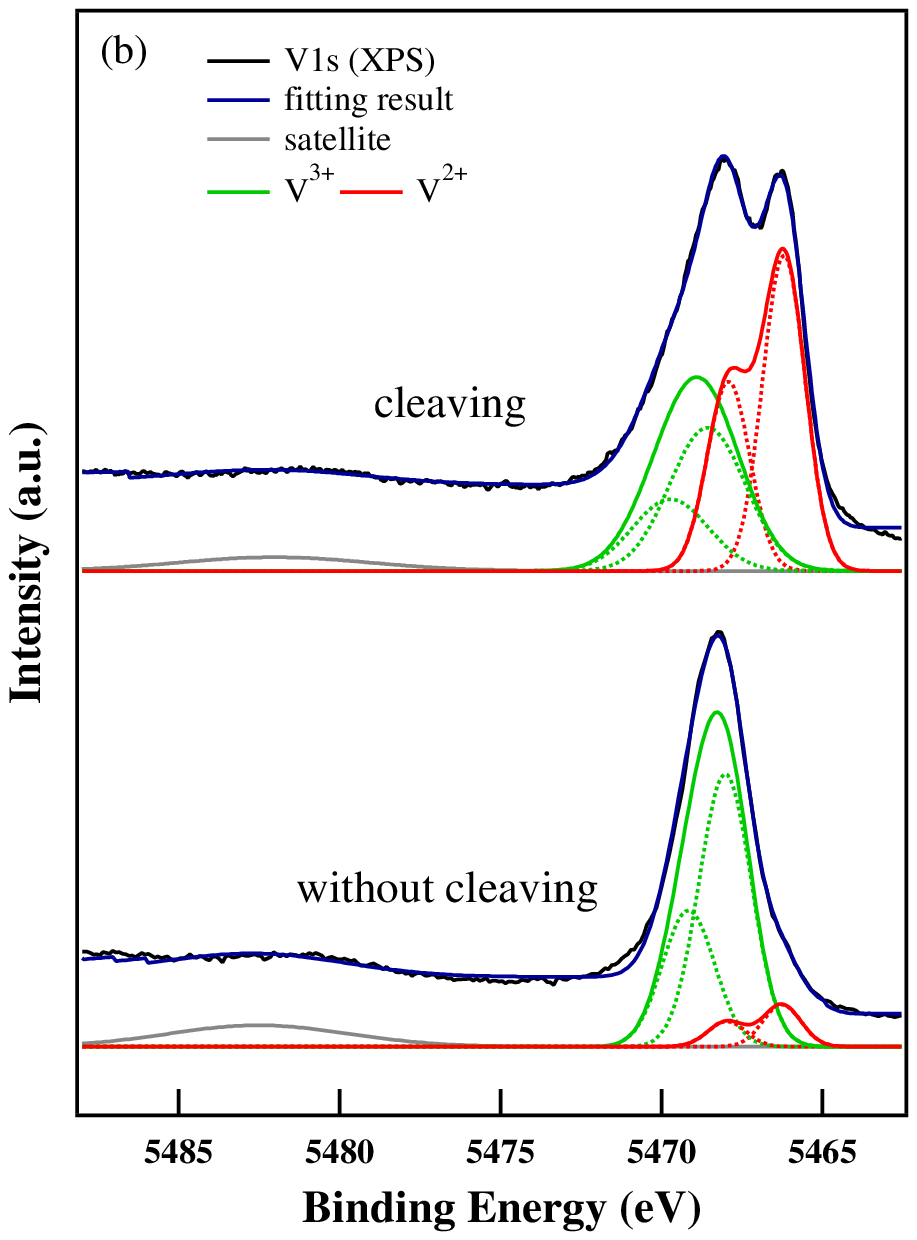}
\caption{
(Color online) 
(a) V 1$s$ HAXPES spectra for clean and oxidized surfaces. 
(b) Upper panel: Gaussian fitting for the spectra for the clean surface. Lower panel: Gaussian fitting for the spectra for the oxidized surface. The V 1$s$ peaks are decomposed by four gaussians as indicated by the dotted curves.
}
\end{center}
\end{figure}

Figure 3(a) shows V 2$p$ and O 1$s$ HAXPES spectra taken at room temperature for the clean and oxidized surfaces. The V 2$p$ spectrum of oxidized surface is dominated by V$^{3+}$ component with minor contribution of V$^{2+}$ as shown in the lower panel of Fig. 3(b). The O 1$s$ peak is very broad for the oxidized surface. The clean surface provides V 2$p$ spectra with comparable V$^{2+}$ and V$^{3+}$ components. In case of V 2$p$, the effect of Coulomb interaction (multiplet splitting) between the core hole and the V 3$d$ electrons is rather complicated. With two Gaussians for V$^{3+}$ and additional two for V$^{2+}$, the V 2$p_{3/2}$ spectrum can be fitted as shown in the lower panel of Fig. 3(b). Here, the V 2$p_{1/2}$ spectrum is given by a Gaussian. The amount of V$^{3+}$ is overestimated in this analysis probably due to the uncertainty of the multiplet splitting. The energy separation between V$^{2+}$ and V$^{3+}$ for V 2$p$ HAXPES is consistent with that for V 1$s$ HAXPES. The O 1$s$ spectra of the clean surface can be decomposed into the low energy peak from the V$_2$O part, {\it i.e.} the oxygen site bridging the stacked VO$_6$ chains, and the high energy peak from the PO$_4$ group. The area ratio between the two peaks is consistent with the number of oxygen atoms in the V$_2$O part and the PO$_4$ group.

\begin{figure}
\begin{center} 
\includegraphics[width=10cm]{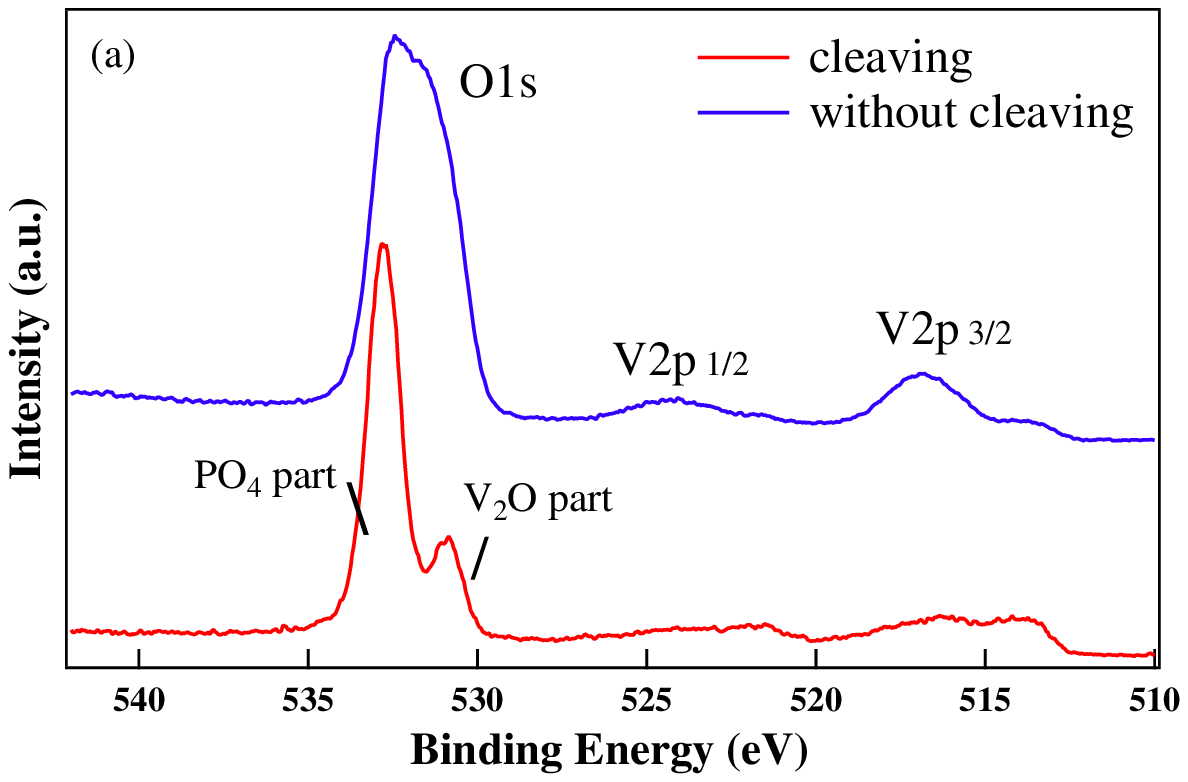}
\includegraphics[width=10cm]{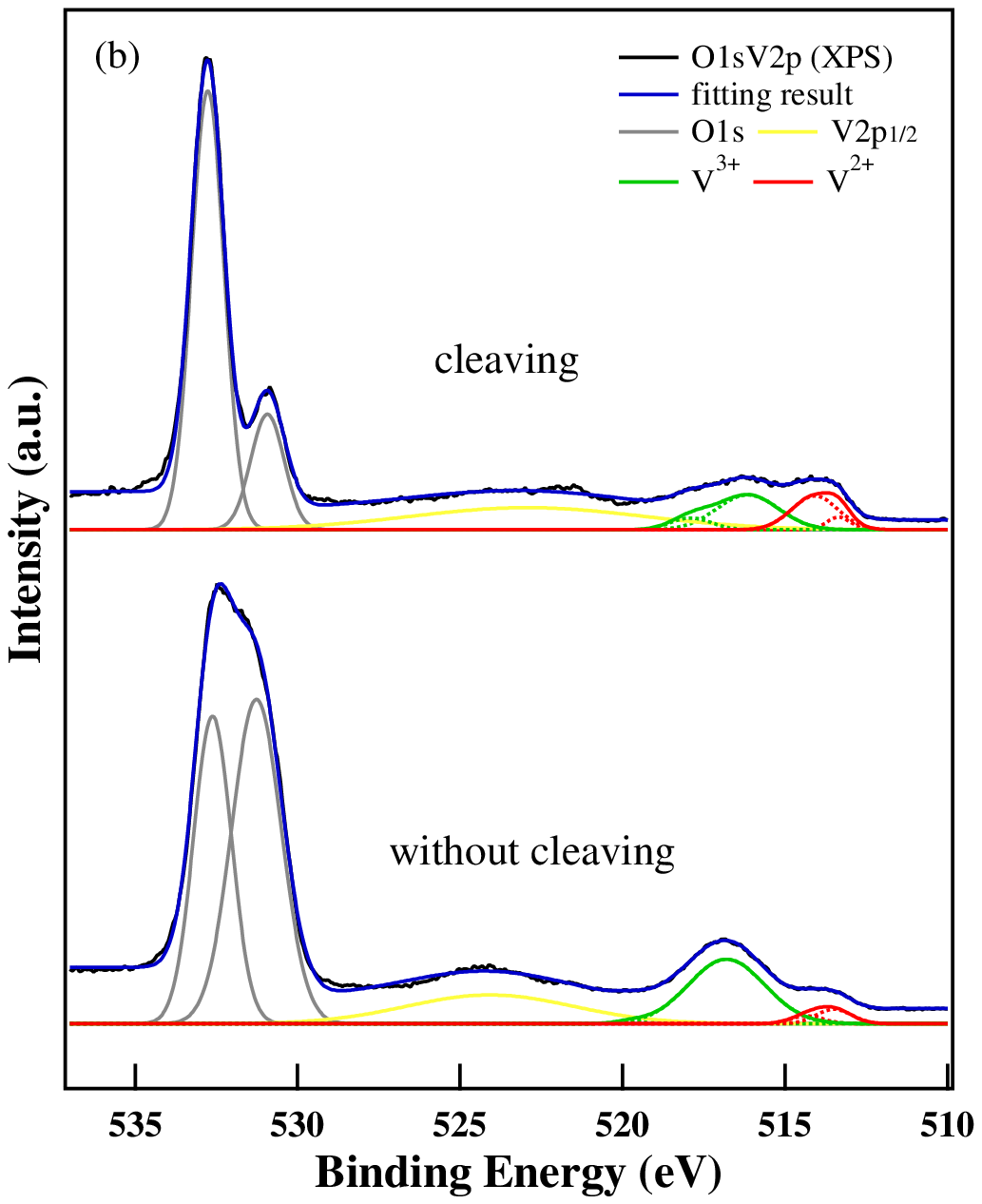}
\caption{
(Color online) 
(a) V 2$p$ and O 1$s$ HAXPES spectra for clean and oxidized surfaces. 
(b) Upper panel: Gaussian fitting for the spectra for the clean surface. Lower panel: Gaussian fitting for the spectra for the oxidized surface. The V 2$p_{3/2}$ and O 1$s$ peaks are decomposed by four gaussians (indicated by the dotted curves) and two gaussians, respectively. The broad V 2$p_{1/2}$ peak is fitted by one gaussian.
}
\end{center}
\end{figure}

\begin{figure}
\begin{center} 
\includegraphics[width=10cm]{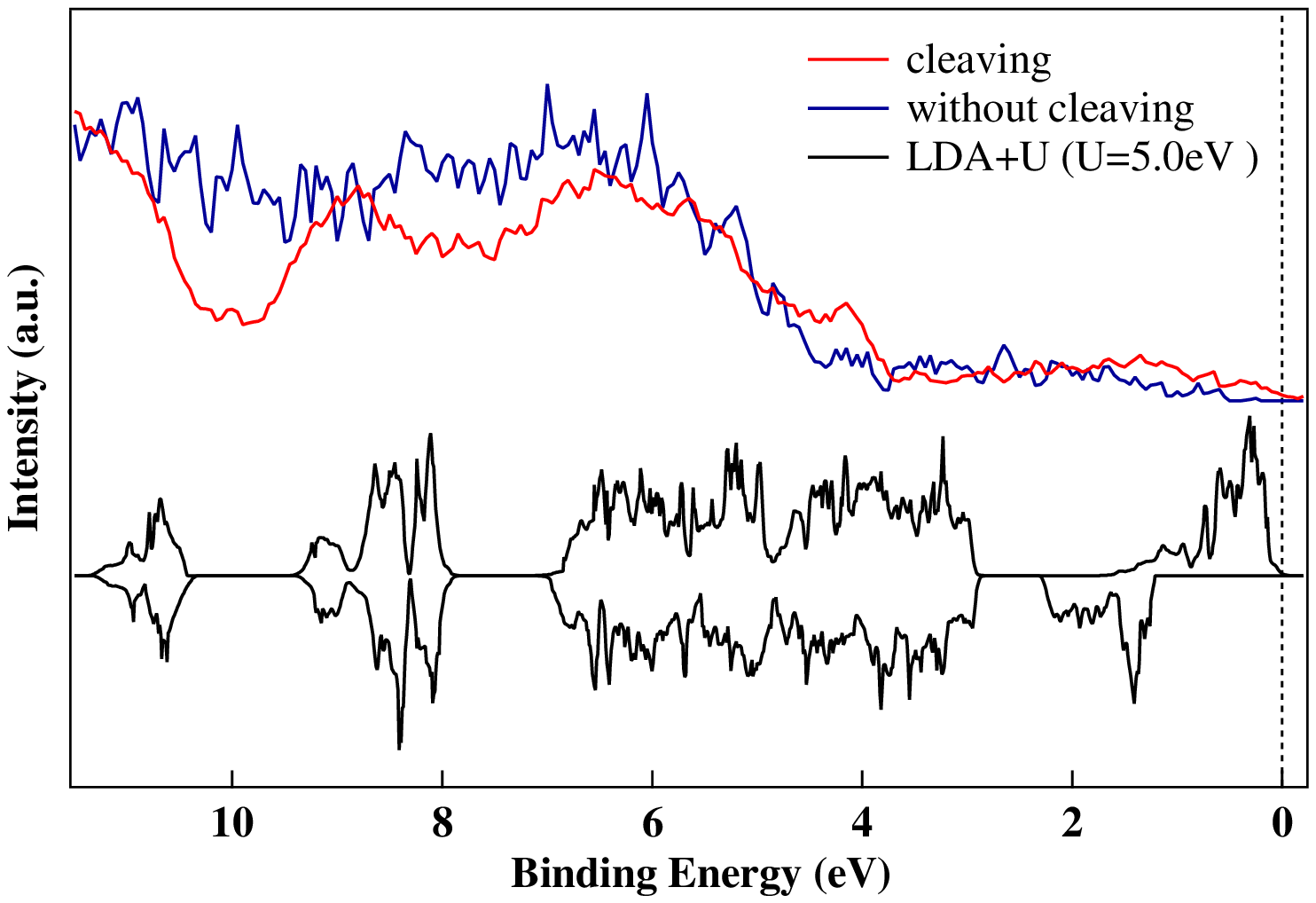}
\caption{
(Color online) 
Valence band HAXPES spectra for clean and oxidized surfaces. The spectra are compared with density of states obtained by LDA+$U$ calculation with $U$ = 5 eV. In the density of states, the upper and lower components correspond to majority and minority spin bands. Near the Fermi level, the majority spin band is dominated by $t_{2g}$ orbitals of the V$^{2+}$ site and the minority spin band is by those of the V$^{3+}$ site.
}
\end{center}
\end{figure}

Figure 4 shows valence band spectra compared with density of states calculated by LDA+$U$ with $U$=5.0 eV and $J$=0.5 eV. 
The valence band spectrum of the oxidized surface is very broad and is considerably different from the calculation. In the clean surface, the O 2$p$ and P 3$p$ bands ranging from 4 eV to 10 eV exhibit several structures which are consistent with the calculation except the shift to the higher binding energy. In the oxidized surface dominated by V$^{3+}$,  the V 3$d$ peak is located around 2 eV below the Fermi level. As for the clean surface, the V 3$d$ spectral weight near the Fermi level is enhanced. Therefore, the V 3$d$ spectral weight near the Fermi level can be assigned to the $t_{2g}$ electrons of the V$^{2+}$ site. Indeed, this assignment is consistent with the calculation in which the density of states near the Fermi level is given by majority spin band from V$^{2+}$. The minority spin band from V$^{3+}$ is located around 1.5 eV below the Fermi level in the calculation which is roughly consistent with the HAXPES spectrum for the clean surface. However, the V 3$d$ spectral distribution is rather broad in the experimental result and is not separated into V$^{2+}$ and V$^{3+}$ bands. Also the spectral weight near the Fermi level is rather large compared to the metallic phase of BaV$_{10}$O$_{15}$ \cite{Dash2019} and Ba$_{1-x}$Sr$_x$V$_{13}$O$_{18}$ \cite{Dash2017}. Here, we speculate that the V$^{2+}$ and V$^{3+}$ components are highly mixed due to strong charge fluctuation along the face sharing bond. Such charge fluctuation by the electron correlation is beyond the mean field treatment of LDA+$U$. The strong hybridization between V$^{2+}$ and V$^{3+}$ and the charge transfer from the V$^{2+}$ site to the V$^{3+}$ site are consistent with the reduced magnetic moment in the paramagnetic phase \cite{Pachoud2018}.
Figure 5 shows density of states calculated by LDA, LDA+$U$ (with $U$=3.5 and 5.0 eV), and GGA+$U$ (with $U$=5.0 eV).
The V 3$d$ $t_{2g}$ density of states at the Fermi level is high in the LDA result. The magnitude of band gap increases with $U$ in the LDA+$U$ calculations. In the GGA+$U$ calculation with $U$=5.0 eV, the spectral shape of the occupied V 3$d$ $t_{2g}$ band is very similar to that obtained by the LDA+$U$ calculation although the magnitude of band gap increases by about 0.2 eV. The valence band shape and the energy difference between the occupied V$^{2+}$ and V$^{3+}$ bands are mainly determined by the charge transfer effect between the V$^{2+}$ and V$^{3+}$ sites and is less sensitive to $U$. On the other hand, the magnitude of band gap is sensitive to $U$ since the energy difference between the occupied and unoccupied V$^{3+}$ bands is controlled by $U$.

\begin{figure}
\begin{center} 
\includegraphics[width=10cm]{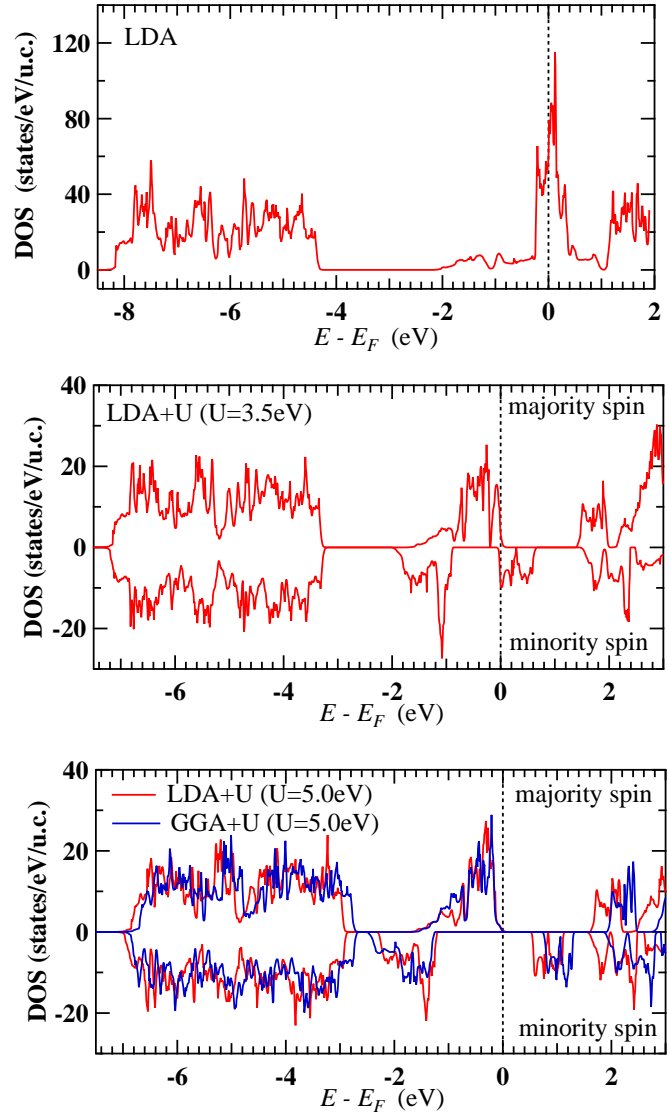}
\caption{
(Color online) 
Density of states calculated by LDA, LDA+$U$ with $U$=3.5 eV, LDA+$U$ and GGA+$U$ with $U$=5.0 eV. The LDA+$U$ and GGA+$U$ calculations are performed for the ferrimagnetic state and the density of states are plotted for the majority spin and minority spin components. $E_F$ represents the Fermi level for the metallic states and the valence band top for the insulating states.
}
\end{center}
\end{figure}

In order to examine the effect of charge transfer between the neighboring V sites, we have analyzed the hybridization between the V 3$d$ and O 2$p$ orbitals in the V$_2$O$_9$ cluster which is illustrated in Fig. 6(a). Molecular orbitals are constructed from the O 2$p_x$, O 2$p_y$, and O 2$p_z$ orbitals of the nine oxygen sites and their energy levels are shown in Fig. 6(b). Among the 27 molecular orbitals constructed from the O 2$p$ orbitals, 6 molecular orbitals exclusively hybridize with the V 3$d$ $a_{1g}$ ($3z^2-r^2$) orbitals at the V sites. Three of them are even with respective to the horizontal mirror plane of the cluster (labeled as Z), and the other three are odd (labeled as Z*). 18 molecular orbitals are doubly degenerate and hybridize with the V 3$d$ $e_g^{\pi}$ and $e_g$ ($x^2-y^2$, $xy$, $zx$, and $yz$) orbitals. Ten of them are even (labeled as XY) and eight of them are odd (labeled as XY*) with respective to the horizontal mirror plane. The vertical mirror plane plays a role to set the base orbitals to $1/\sqrt{3}(x^2-y^2)-\sqrt{2/3}zx$ and $1/\sqrt{3}xy-\sqrt{2/3}yz$ for $e_g^{\pi}$, $\sqrt{2/3} (x^2-y^2)+ 1/\sqrt{3} zx$ and $\sqrt{2/3}xy+1/\sqrt{3}yz$ for $e_g$ \cite{Kugel2015}. In addition, there are three non-bonding orbitals (labeled as NB). Under the cubic symmetry, the ground state of V$^{3+}$ ($d^2$) is $^3$T$_1$ and that of V$^{2+}$ ($d^3$) is $^4$A$_2$. The trigonal symmetry of the V$_2$O$_9$ split $^3$T$_1$ into $^3$A$_1$ and $^3$E. In the present calculation, the orbitally degenerate ground state is obtained by a combination of $^3$E for V$^{3+}$ ($d^2$) and $^4$A$_2$ for V$^{2+}$ ($d^3$). This orbital degeneracy is consistent with the orbital order reported in the recent x-ray absorption study \cite{Murota2020}. The distortion of the V$_2$O$_9$ cluster is modelled as illustrated in Fig. 6(a). The VO$_6$ octahedron for the V$^{2+}$ site (V$_2$ site) is elongated along the $z$ axis as indicated by the arrows in the side view. The distortion of the shared triangle is represented by the shift of oxygen (O$_4$ site) along the $y$ axis as indicated by the arrows in the top view of Fig. 6(a). This distortion breaks the symmetry due to the vertical mirror plane ($xz$ plane in Fig. 6(a)) and the trigonal symmetry. As a result, the degeneracy of the molecular orbitals is lifted as indicated in Fig. 6(b). In addition, the orbital degeneracy of the ground state is lifted.
 
\begin{figure}
\begin{center} 
\includegraphics[width=10cm]{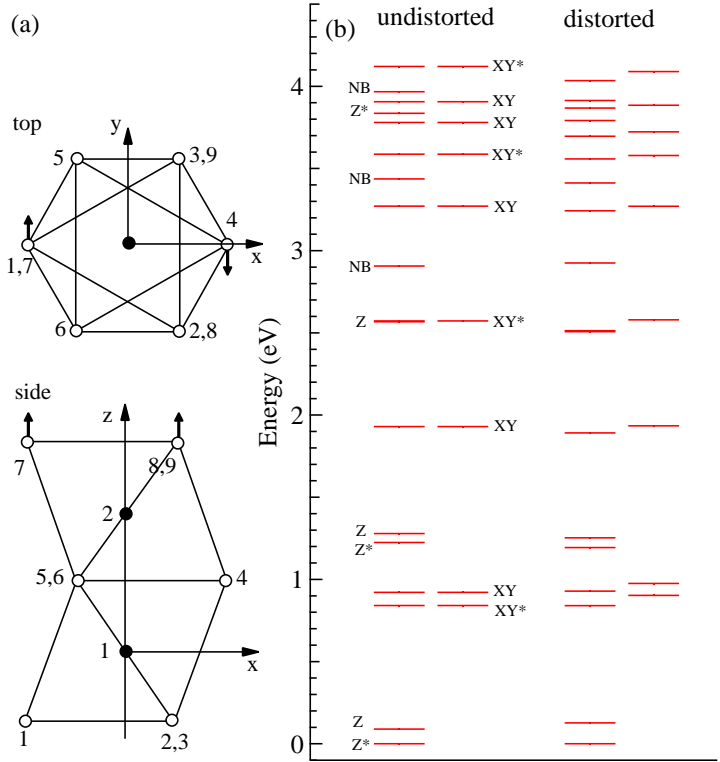}
\caption{
(Color online) 
(a)V$_2$O$_9$ cluster model with two VO$_6$ octahedra sharing their face. The $z$ axis is perpendicular to the shared face. (b) Energy levels of O 2$p$ molecular orbitals in the V$_2$O$_9$ cluster with and without the distortion.
}
\end{center}
\end{figure}

The ground state is sensitive to the energy difference between the $d_1^2d_2^3$ and $d_1^3d_2^2$ ($E_G$). The relationship between $E_G$ and the upper and lower Hubbard bands for the V$^{3+}$ and V$^{2+}$ sites is illustrated in Fig. 7(a). From the LDA+$U$ calculations \cite{Pachoud2018, Murota2020}, $E_G$ is deduced to be $\sim$ 1.5 eV.  Considering the charge transfer between the two V sites and that from the O 2$p$ molecular orbitals to the V 3$d$ orbitals, we have calculated the V 1$s$ photoemission spectra as shown in Fig. 7(b). With $E_G$ = 0.0 eV, the $d_1^2d_2^3$ and $d_1^3d_2^2$ configurations are almost equally hybridized in the ground state. When $U_c$ is about 2.0 eV or smaller, the $cd_1^2d_2^3$ and $cd_1^3d_2^2$ configurations are almost equally hybridized in the final states. Here, $c$ represents a V 1$s$ core hole at the $d_1$ site. Then the transition to the antibonding state is almost suppressed disagreeing with the experimental result. With $U_c$ of 3.0 eV, the bonding state has more $cd_1^3d_1^2$ character corresponding to V$^{2+}$ final state while the antibonding state has $cd_1^2d_1^3$ character corresponding to V$^{3+}$ final state. When $U_c$ is about 4.0 eV or larger, the splitting of the V$^{2+}$ and V$^{3+}$ components becomes too large compared to the experimental result. This situation is improved by including $E_G$.

With $E_G$ = 1.5 eV, the splitting of the V$^{2+}$ and V$^{3+}$ components is roughly reproduced by the calculation with ($pd\sigma$) = -1.5 eV, $U_c$ = 4.0 eV, and $\Delta$ = 6.0 eV. However, with ($pd\sigma$) = -1.5 eV, the position of the charge transfer satellite is too close to the main peak compared to the experimental result. The agreement for the satellite structure is improved by increasing the magnitude of ($pd\sigma$). At the same time, the splitting between V$^{2+}$ and V$^{3+}$ components increases and the intensity of V$^{3+}$ component decreases with ($pd\sigma$) which is responsible for the hybridization between the two V sites via the O 2$p$ orbitals. In order to reproduce both the charge transfer satellite and the V$^{2+}$-V$^{3+}$ splitting, charge transfer energy $\Delta$ should be reduced from 6.0 eV to 4.0 eV with ($pd\sigma$) = -2.0 eV. Since $\Delta$ of 4.0 eV is too small for V$^{3+}$ or V$^{2+}$ oxides, we speculate that the intersite interactions between the V 1$s$ core hole and the neighboring V 3$d$ electrons should be taken into account to refine the parameter values. On the other hand, inclusion of the intersite Coulomb interaction makes the cluster model more complicated than the standard models which have been applied to various transition-metal oxides. The energy splitting between the ground state and the excited state of the V$_2$O$_9$ cluster also increases with increasing the hybridization. The position of the lower Hubbard band of the V$^{2+}$ site is expected to get closer to that from the V$^{3+}$ site since the charge difference is reduced by the hybridization between the two sites. This situation is consistent with the valence band spectra.

\begin{figure}
\begin{center} 
\includegraphics[width=10cm]{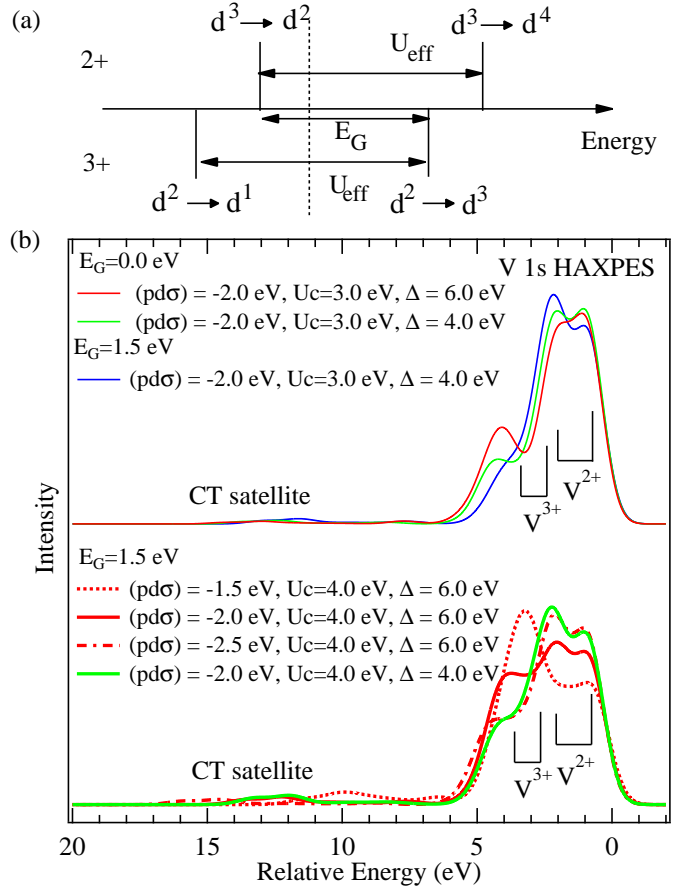}
\caption{
(Color online) 
(a) Schematic picture of the lower and upper Hubbard bands for the  V$^{2+}$ and V$^{3+}$ sites. $U_{eff}$ represents the effective on-site Coulomb interaction which is reduced from $U$ due to hybridization with the O 2$p$ orbitals. 
(b)V 1$s$ HAXPES spectra calculated by the V$_2$O$_9$ cluster model. 
($pd\sigma$) is the transfer integral between the V 3$d$ and O 2$p$ orbitals. $U_c$ is the Coulomb interaction between the V 1$s$ and V 3$d$ electron. 
$\Delta$ represents the O 2$p$ to V 3$d$ charge transfer energy for V$^{3+}$. 
}
\end{center}
\end{figure}

\section{Conclusion}
In conclusion, the electronic structure of mixed valence V$_2$OPO$_4$ has been investigated by means of hard x-ray photoemission spectroscopy and subsequent theoretical calculations. The V 1$s$ and 2$p$ HAXPES spectra are consistent with charge ordering of V$^{2+}$ : V$^{3+}$ = 1: 1. The binding energy difference between the V$^{2+}$ and V$^{3+}$ components is unexpectedly large indicating strong hybridization between them in the final states of core level photoemission. The valence band top is given by the V 3$d$ $t_{2g}$ orbitals of the V$^{2+}$ site which is followed by those of the V$^{3+}$ site. The broad V 3$d$ spectral distribution is consistent with the strong hybridization and the charge transfer from the V$^{2+}$ site to the V$^{3+}$ site indicating that the charge ordering on V$_2$OPO$_4$ is not complete.

\section*{Acknowledgements}
The work at Waseda was supported by CREST-JST (Grant No. JPMJCR15Q2), KAKENHI from JSPS (Grants No.19H01853 and No.19H00659). The work at Edinburgh was supported by EPSRC and ERC. The work of D.I. Khomskii was funded by the Deutsche Forschungsgemeinschaft (DFG, German Research Foundation) - Project number 277146847 - CRC 1238. The synchrotron radiation experiment was performed with the approval of SPring-8 (2019B1574).

\end{document}